\documentstyle[aps,manuscript,epsfig]{revtex}

\begin{document}

\title{Effective noise in stochastic description of inflation}

\author{S. Winitzki}

\address{DAMTP, University of Cambridge, Cambridge CB3 9EW, UK\thanks{
Address after Nov. 1, 1999: Department of Physics, Case Western Reserve University,
10900 Euclid Avenue, Cleveland, OH 44106-7079, USA.
}}

\author{A. Vilenkin}

\address{Tufts Institute of Cosmology, Department of Physics and Astronomy, Tufts University,
Medford MA 02155, USA.}

\maketitle
\begin{abstract}
Stochastic description of inflationary spacetimes emulates the growth of vacuum
fluctuations by an effective stochastic ``noise field'' which drives the dynamics
of the volume-smoothed inflaton. We investigate statistical properties of this
field and find its correlator to be a function of distance measured in units
of the smoothing length. Our results apply for a wide class of smoothing window
functions and are different from previous calculations by Starobinsky and others
who used a sharp momentum cutoff. We also discuss the applicability of some
approximate noise descriptions to simulations of stochastic inflation.
\end{abstract}

\draft

\section{Introduction}

Inflationary models explain structure formation as a consequence of vacuum fluctuations
of a scalar field \( \phi  \), the inflaton \cite{Inflation}. Quantum fluctuations
are also responsible for the structure of the universe at super-large scales
and for the eternal character of inflation \cite{Vil83,Linde86,Starob1}. Because
of exponentially rapid expansion of the spacetime, fluctuations of the inflaton
field \( \phi  \) on super-horizon scales effectively ``freeze'' a few Hubble
times \( H^{-1} \) after they leave the horizon and can be treated as contributions
to classical values of the inflaton field. After smoothing on super-horizon
scales, the averaged field \( \Phi  \) can be considered to be classical with
an approximate equation of motion driven by the inflaton potential \( V\left( \Phi \right)  \)
and by small-scale ``noise'' \( \xi  \), 
\begin{equation}
\label{eq:eqmot-sr-xi}
\dot{\Phi }=-\frac{1}{3H}V^{\prime }\left( \Phi \right) +\xi \left( {\mathbf x},t\right) .
\end{equation}
 Here the ``noise field'' \( \xi \left( {\mathbf x},t\right)  \) is a Gaussian
random field characterized by its two-point function \( \left\langle \xi \left( {\mathbf x},t\right) \xi \left( {\mathbf x}^{\prime },t^{\prime }\right) \right\rangle  \).
It is important to know the correct behavior of this correlator, in particular
for numerical simulations of eternal inflation.

One can approximate the noise field by using Eq.~(\ref{eq:eqmot-sr-xi}) with
a free scalar field in de~Sitter spacetime, even though the scalar field is
only approximately massless and the spacetime is approximately de~Sitter. The
results will be applicable on scales smaller than the characteristic scale of
the variation of the expansion rate \( H \) and that scale is typically exponentially
large. In the pioneering work of Starobinsky \cite{Starob1} the correlator
was evaluated in this manner as
\begin{equation}
\label{eq:starob-corr}
\left\langle \xi \left( {\mathbf x},t\right) \xi \left( {\mathbf x}^{\prime },t^{\prime }\right) \right\rangle =\left( \frac{H}{2\pi }\right) ^{2}\frac{\sin \epsilon He^{Ht}\left| {\mathbf x}-{\mathbf x}^{\prime }\right| }{\epsilon He^{Ht}\left| {\mathbf x}-{\mathbf x}^{\prime }\right| }\delta \left( t-t^{\prime }\right) .
\end{equation}
The smoothing scale for the inflaton field was \( R\equiv \left[ \epsilon H\exp \left( -Ht\right) \right] ^{-1} \)
with \( \epsilon \ll 1 \). Eq.~(\ref{eq:starob-corr}) shows a surprisingly
slow decay of correlations at large distances. For comparison, the two-point
function of the time derivatives of the un-smoothed field \( \left\langle \dot{\phi }\left( {\mathbf x},t\right) \dot{\phi }\left( {\mathbf x}^{\prime },t^{\prime }\right) \right\rangle  \)
at large separations \( r\equiv \left| {\mathbf x}-{\mathbf x}^{\prime }\right|  \)
behaves as \( \propto r^{-4} \) (here the angular brackets denote vacuum expectation
value rather than statistical average). One would not expect a smearing of the
field operators \( \phi \left( {\mathbf x},t\right)  \) on scales \( R \) to
have such an effect on correlations at distances \( r\gg R \).

Our analysis shows that the origin of the unusual behavior of the correlator
found by Starobinsky is the sharp momentum cutoff in his smoothing procedure.
With a smooth cutoff, we recover the \( r^{-4} \) behavior independently of
the cutoff window function and find that the time dependence of the noise correlator
at large times is generically \( \propto \exp \left( -2Ht\right)  \) instead
of a sharp \( \delta  \)-function dependence of Eq.~(\ref{eq:starob-corr}).

The correlator of noise is also important for simulations of the inflating spacetime
using Eq.~(\ref{eq:eqmot-sr-xi}). One of the simulation methods introduced
in Ref.~\cite{LLMlarge} consists of approximating Eq.~(\ref{eq:eqmot-sr-xi})
by a finite difference equation with random sine waves playing the role of \( \xi  \),
\begin{equation}
\label{eq:sw-appr}
\phi \left( t+\Delta t,{\mathbf x}\right) -\phi \left( t,{\mathbf x}\right) =-\frac{V^{\prime }\left( \phi \right) \Delta t}{3H}+A\sin \left( He^{Ht}{\mathbf nx}+\alpha \right) .
\end{equation}
Here \( {\mathbf n} \) is a randomly directed unit vector, \( \alpha  \) is
a random phase and \( A \) is an appropriately distributed random amplitude.
We investigate the limits of validity of this approximation and show that for
any finite \( \Delta t \), the long-distance correlation of the effective noise
is asymptotically the same as that given by Eq.~(\ref{eq:starob-corr}) which,
as we argue, is unphysically large. Even with \( \Delta t\rightarrow 0 \),
the sum of independent sine waves over one Hubble time \( H^{-1} \) produces
an effective noise field with correlations \( \propto r^{-2} \). This method
therefore can only be used in contexts that do not require precise simulation
of the large-scale correlations of noise. In cases where such correlations are
important, the noise field may be simulated as a Gaussian random field with
known correlator; this method was employed in Ref.~\cite{VVW99}.

The paper is organized as follows. The results regarding the noise correlator
are presented in Section \ref{sec:corr-sec}, and the analysis of the sine wave
approximation of Eq.~(\ref{eq:sw-appr}) is in Section~\ref{sec:sinwaves}.
The necessary details of the calculations are delegated to appendices. Appendix~\ref{sec:app-w}
describes the properties of window functions used for spatial smoothing, Appendix~\ref{sec:app-corr}
contains a derivation of the noise correlator with arbitrary window, and Appendix~\ref{sec:app-desitter}
lists the correlators of unsmoothed field in de~Sitter space for reference purposes.

\section{Spatial averaging and noise}

\label{sec:corr-sec}Field fluctuations on super-horizon scales behave effectively
as classical fluctuation modes with random amplitudes. This is conventionally
described \cite{Starob1,Starob2,Nambu,Mijic,Salopek} by averaging the field
\( \phi  \) in space over super-horizon scales and treating the resulting field
\( \bar{\phi } \) as a classical (albeit stochastic) field \( \Phi  \) satisfying
an equation of motion with a stochastic term (or ``noise field''). This term
is a Gaussian random field which describes the effective character of the field
fluctuations; it will be the focus of this section.

We consider a free massless scalar field \( \phi  \) in an inflating (de~Sitter)
spacetime with horizon size \( H^{-1} \) and the scale factor \( a\left( t\right) =\exp \left( Ht\right)  \).
The equation of motion for the free field \( \phi  \) is
\begin{equation}
\label{eq:eqmot}
\Box \phi =\ddot{\phi }+3H\dot{\phi }-\frac{1}{a\left( t\right) ^{2}}\Delta \phi =0.
\end{equation}
 The field is quantized via the usual mode expansion (see Eqs.~(\ref{eq:phime}),
(\ref{eq:phimf}) of Appendix~\ref{sec:app-corr}),
\begin{equation}
\phi \left( {\mathbf x},t\right) =\int \frac{d^{3}{\mathbf k}}{\left( 2\pi \right) ^{3/2}}\left( a_{{\mathbf k}}\psi _{k}\left( t\right) e^{i{\mathbf kx}}+h.c.\right) .
\end{equation}
The averaging of the field \( \phi  \) is performed by means of a suitable
smoothing window \( W_{s}\left( {\mathbf x};R\right)  \) with a characteristic
smoothing scale \( R \),
\begin{equation}
\bar{\phi }\left( {\mathbf x},t\right) \equiv \int d^{3}{\mathbf x}^{\prime }\phi \left( {\mathbf x}^{\prime },t\right) W_{s}\left( {\mathbf x}-{\mathbf x}^{\prime };R\right) .
\end{equation}
Here, the physical smoothing scale is taken to be \( \epsilon ^{-1} \) times
larger than the horizon size, with \( \epsilon \ll 1 \). The corresponding
comoving scale is 
\begin{equation}
R\equiv R\left( t\right) =\frac{1}{\epsilon Ha\left( t\right) }.
\end{equation}
The volume-averaged field has a mode expansion given by Eq.~(\ref{eq:sm-me}),
\begin{equation}
\bar{\phi }\left( {\mathbf x},t\right) =\int \frac{d^{3}{\mathbf k}}{\left( 2\pi \right) ^{3/2}}w\left( kR\right) a_{{\mathbf k}}\psi _{k}\left( t\right) e^{i{\mathbf kx}}+h.c.,
\end{equation}
where \( w\left( kR\right)  \) is a suitable Fourier transform of the window
profile \( W_{s} \) (see Eqs.~(\ref{eq:wprof}), (\ref{eq:g-def})).

The volume-averaged inflaton field is treated as a classical field \( \Phi  \)
satisfying Eq.~(\ref{eq:eqmot-sr-xi}) which is a Langevin equation describing
dynamics driven by an effective ``noise field'' \( \xi \left( {\mathbf x},t\right)  \)
as well as by the effective potential \( V\left( \Phi \right)  \). The noise
field \( \xi \left( {\mathbf x},t\right)  \) in that equation can be heuristically
defined as a stochastic field that corresponds to the quantum operator of the
free field derivative \( \dot{\bar{\phi }} \), in the sense that any averages
of \( \xi  \), such as the correlator \( \left\langle \xi \left( {\mathbf x},t\right) \xi \left( {\mathbf x}^{\prime },t^{\prime }\right) \right\rangle  \),
are taken to be the same as the corresponding quantum expectation values of
\( \dot{\bar{\phi }} \) in the vacuum state (the standard Bunch-Davies vacuum).
The effective noise field \( \xi  \) defined in this way is a Gaussian random
field with zero mean, so the correlator \( \left\langle \xi \left( {\mathbf x},t\right) \xi \left( {\mathbf x}^{\prime },t^{\prime }\right) \right\rangle  \)
completely describes its properties. The calculation of the noise correlator
\( \left\langle \xi \left( {\mathbf x},t\right) \xi \left( {\mathbf x}^{\prime },t^{\prime }\right) \right\rangle  \)
given in Appendix~\ref{sec:app-corr} is a straightforward computation of the
corresponding expectation value of the quantum ``noise operator'' \( \dot{\bar{\phi }} \).

The noise correlator generally depends on the particular window function \( W_{s}\left( {\mathbf x};R\right)  \)
and on the parameter \( \epsilon  \). Since the smoothing window is merely
a technical device in this approach, we would expect to obtain results independent
of the window \( W_{s}\left( {\mathbf x};R\right)  \) on scales \( \gg R \)
for a reasonably wide class of window shapes. It seems natural to require that
the window function should be non-negative, spherically symmetric and depend
on \textbf{\( {\mathbf x} \)} only through the combination \( \left| {\mathbf x}\right| /R \).
The last requirement makes \( W_{s} \) a function of the form \( W_{s}\left( {\mathbf x};R\right) =R^{-3}W\left( \left| {\mathbf x}\right| /R\right)  \)
with some suitable profile function \( W\left( q\right)  \) which starts to
decrease rapidly at \( q\sim 1 \) (see Eqs.~(\ref{eq:wprof})--(\ref{eq:norm-cond})).
Additionally, we require that the window profile decays rapidly enough at large
distances, as detailed below.

It turns out, as shown in Appendix~\ref{sec:app-corr}, that the asymptotics
of the noise correlator at large temporal or spatial separations are independent
of the choice of the window profile \( W\left( r\right)  \) if it satisfies
these conditions. The positivity assumption, \( W\left( r\right) \geq 0 \),
may in fact be relaxed and substituted by the condition 
\begin{equation}
\label{eq:cond-w}
\left\langle r^{2}\right\rangle _{W}\equiv 4\pi \int _{0}^{\infty }W\left( r\right) r^{4}dr>0.
\end{equation}
For the integral in Eq.~(\ref{eq:cond-w}) to converge, we require that the
window profile decays at large distances as \( W\left( r\right) \sim r^{-6} \)
or faster.

The noise correlator is a function of distance and of the smoothing scale \( \epsilon  \),

\begin{equation}
\label{eq:D-def}
\left\langle \xi \left( {\mathbf x},t=0\right) \xi \left( {\mathbf x}^{\prime },t\right) \right\rangle \equiv C\left( \left| {\mathbf x}-{\mathbf x}^{\prime }\right| ,t;\, \epsilon \right) .
\end{equation}
While the precise shape of this function does depend on the window profile \( W\left( r\right)  \),
the large-distance asymptotic does not (cf.~Eq.~(\ref{eq:smcor-asy})):
\begin{equation}
\label{eq:d-asymp}
C\left( \left| {\mathbf x}-{\mathbf x}^{\prime }\right| ,t;\, \epsilon \right) \propto e^{-2Ht}\left( \epsilon H\left| {\mathbf x}-{\mathbf x}^{\prime }\right| \right) ^{-4}+O\left( \left| {\mathbf x}-{\mathbf x}^{\prime }\right| ^{-6}\right) .
\end{equation}
 This behavior agrees with the \( r^{-4} \) asymptotic behavior of the field
derivatives in de Sitter space (see Eq.~(\ref{eq:shcor-asy})). However, Eq.~(\ref{eq:d-asymp})
is in disagreement with the result of Starobinsky \cite{Starob1} where the
correlator was found to have a much higher correlation of noise at large distances
(Eq.~(\ref{eq:starob-corr})).

The discrepancy between Eqs.~(\ref{eq:starob-corr}) and (\ref{eq:d-asymp})
is due to different choices of the smoothing window profiles. Ref.~\cite{Starob1}
used a sharp step-function cutoff in Fourier space,

\begin{equation}
w\left( kR\right) =\theta \left( 1-kR\right) 
\end{equation}
which corresponds to the real-space smoothing window

\begin{equation}
\label{eq:starob-w}
W\left( x\right) =\frac{\sin x-x\cos x}{2\pi ^{2}x^{3}}.
\end{equation}
The smoothing window of Eq.~(\ref{eq:starob-w}) is not everywhere positive
and decays too slowly at large distances to satisfy Eq.~(\ref{eq:cond-w}).
The latter condition is equivalent to the requirement that \( w\left( kR\right)  \)
be a sufficiently smooth function (at least twice differentiable). If we allow
windows \( W\left( x\right)  \) with discontinuous Fourier transforms \( w\left( kR\right)  \),
then different choices of window would lead to a wide range of asymptotic behaviors
of the correlator. On the other hand, as long as Eq.~(\ref{eq:cond-w}) is satisfied,
the asymptotic Eq.~(\ref{eq:d-asymp}) is independent of the choice of the window
shape \( w\left( kR\right)  \) or equivalently \( W\left( x\right)  \). Therefore
we conclude that the long-distance behavior of the correlator obtained in Ref.~\cite{Starob1}
is an artefact of the sharp mode cutoff.

In the limit of small \( \epsilon  \) we find that the noise correlator is
a function only of \( \rho \equiv \epsilon H\left| {\mathbf x}-{\mathbf x}^{\prime }\right|  \)
(the distance measured in smoothing scale units) and \( t \) (see Eq.~(\ref{eq:smcorgen})).
It means that the correlator has a scaling property,
\begin{equation}
\label{eq:D-scale}
C\left( \left| {\mathbf x}-{\mathbf x}^{\prime }\right| ,t;\, \epsilon \Lambda \right) =C\left( \Lambda \left| {\mathbf x}-{\mathbf x}^{\prime }\right| ,t;\, \epsilon \right) +O\left( \epsilon ^{2}\right) .
\end{equation}

\begin{figure}
{\par\centering \epsfig{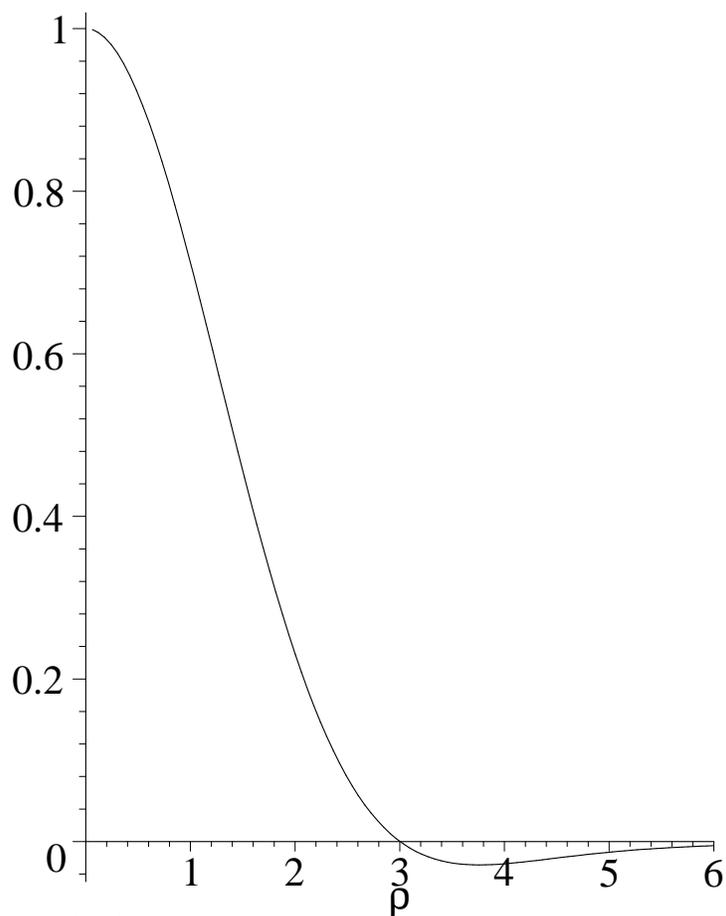} \par}

\caption{\label{fig:corr-shape-a}Correlator \protect\( C\left( \rho ,t\right) \protect \)
of the effective noise field as function of \protect\( \rho \protect \) computed
at equal times (\protect\( t=0\protect \)). The distance was measured in smoothing
scale units (\protect\( \rho \equiv \epsilon Hr\protect \)). }
\end{figure}

For illustration we give some specific results with a Gaussian smoothing window.
The correlator profile in that case is given by Eq.~(\ref{eq:smcor-gau}). The
shape of the correlator is illustrated in Fig.~\ref{fig:corr-shape-a}.

\section{On simulations of the stochastic dynamics}

\label{sec:sinwaves}Simulations of inflating spacetimes use a discretized version
of Eq.~(\ref{eq:eqmot}). One then needs to represent the effective noise field
\( \xi  \) by some random process. An approximation used in Refs.~\cite{LLMlarge,VVW99}
employed random sine waves in lieu of \( \xi  \),
\begin{equation}
\label{eq:xi-sinewave}
\tilde{\xi }\left( {\mathbf x},t\right) =A\sin \left( Ha\left( t\right) {\mathbf nx}+\alpha \right) .
\end{equation}
Here \( {\mathbf n} \) is a randomly directed unit vector, \( \alpha  \) is
a random phase and \( A \) is an appropriately distributed random amplitude.
The equation of motion was discretized using the slow roll approximation Eq.~(\ref{eq:eqmot-sr-xi}),
yielding
\begin{equation}
\label{eq:eqmot-dis}
\phi \left( t+\Delta t,{\mathbf x}\right) -\phi \left( t,{\mathbf x}\right) =-\frac{V^{\prime }\left( \phi \right) \Delta t}{3H}+\tilde{\xi }\left( {\mathbf x},t\right) .
\end{equation}
The time interval \( \Delta t \) was chosen to be much smaller than the Hubble
time \( H^{-1} \). 

The random field \( \tilde{\xi }\left( {\mathbf x},t\right)  \) defined by Eq.~(\ref{eq:xi-sinewave})
is not actually Gaussian, although the one-point distributions of \( \tilde{\xi }\left( {\mathbf x},t\right)  \)
at any fixed point \( {\mathbf x} \) are Gaussian if the random amplitude \( A \)
is drawn from the standard \( \chi ^{2} \) distribution. We can compute the
correlator by averaging over \( A \), \( \alpha  \) and \( {\mathbf n} \)
and find, up to normalization,
\begin{equation}
\left\langle \tilde{\xi }\left( {\mathbf x},t\right) \tilde{\xi }\left( {\mathbf x}^{\prime },t^{\prime }\right) \right\rangle \propto \frac{\sin H\left| a\left( t\right) {\mathbf x}-a\left( t^{\prime }\right) {\mathbf x}^{\prime }\right| }{H\left| a\left( t\right) {\mathbf x}-a\left( t^{\prime }\right) {\mathbf x}^{\prime }\right| }.
\end{equation}
Note that the correlator at \( t=t^{\prime } \) is of the same form as Eq.~(\ref{eq:starob-corr}),
although there is no sharp time dependence. The field \( \tilde{\xi }\left( {\mathbf x},t\right)  \)
exhibits the same large correlations \( \sim r^{-1} \) at large distances as
the noise field computed with the sharp mode cutoff.

One could hope to improve the approximation by adding many small timesteps \( \Delta t\rightarrow 0 \)
in Eq.~(\ref{eq:eqmot-dis}) while keeping the total time increment constant,
\( N\Delta t\sim H^{-1} \). The net effect is that of adding many independent
sine waves together,
\begin{equation}
\tilde{\xi }\left( {\mathbf x},t;\, N\Delta t\right) \equiv \frac{1}{N}\sum ^{N-1}_{n=0}\tilde{\xi }\left( {\mathbf x},t+n\Delta t\right) \approx \frac{1}{N\Delta t}\int _{t}^{t+N\Delta t}\tilde{\xi }\left( {\mathbf x},\tau \right) d\tau .
\end{equation}
The correlator of the resulting random field \( \tilde{\xi }\left( {\mathbf x},t;\, N\Delta t\right)  \)
is

\begin{equation}
\label{eq:xi-t-N}
\left\langle \tilde{\xi }\left( {\mathbf x},t;\, N\Delta t\right) \tilde{\xi }\left( {\mathbf x}^{\prime },t^{\prime };\, N\Delta t\right) \right\rangle \propto \frac{O\left( 1\right) }{H^{2}\left| a\left( t\right) {\mathbf x}-a\left( t^{\prime }\right) {\mathbf x}^{\prime }\right| ^{2}}.
\end{equation}
(The numerator in the last equation is an oscillating function of order \( 1 \),
the explicit form of which will not be important.) The field \( \tilde{\xi }\left( {\mathbf x},t;\, N\Delta t\right)  \)
is an approximation to an increment of the smoothed field \( \Delta \bar{\phi }\equiv \bar{\phi }\left( {\mathbf x},t+N\Delta t\right) -\bar{\phi }\left( {\mathbf x},t\right)  \),
and the correlator in Eq.~(\ref{eq:xi-t-N}) with \( N\Delta t\sim H^{-1} \)
is to be compared with the corresponding correlator of increments, which under
the same assumptions as Eq.~(\ref{eq:d-asymp}) can be shown to behave at long
distances as
\begin{equation}
\left\langle \Delta \bar{\phi }\left( {\mathbf x},t\right) \Delta \bar{\phi }\left( {\mathbf x}^{\prime },t^{\prime }\right) \right\rangle \propto \left| {\mathbf x}-{\mathbf x}^{\prime }\right| ^{-4}.
\end{equation}
We find that the unphysically large correlations at large distances do not disappear
even in the limit of \( \Delta t\rightarrow 0 \).

We conclude therefore that the approximation method of Eq.~(\ref{eq:xi-sinewave})
can only be used in contexts that do not require precise simulation of the large-scale
correlations of noise.

\section{Acknowledgments}

We thank Pavel Ivanov, Slava Mukhanov and Alexey Starobinsky for useful discussions.
GR/ The work of A.V.\ was supported in part by the National Science Foundation.
S.W.\ was supported by PPARC rolling grant GR/L21488. We acknowledge the hospitality
of the Isaac Newton Institute for Mathematical Sciences where part of this work
was completed. S.W.\ is grateful for the hospitality of Tufts Institute of Cosmology
during the final stages of this project.

\appendix

\section{Properties of window function}

\label{sec:app-w}We first investigate properties of the class of window profiles
we will be using for spatial averages of \( \phi  \). Spherical symmetry and
scaling with \( R \) forces the smoothing window profile \( W_{s}\left( {\mathbf x};R\right)  \)
to depend on \textbf{\( {\mathbf x} \)} only through the combination \( \left| {\mathbf x}\right| /R \),
and from the normalization \( \int W_{s}\left( {\mathbf x};R\right) d^{3}{\mathbf x}=1 \)
it follows that 
\begin{equation}
\label{eq:wprof}
W_{s}\left( {\mathbf x};R\right) =R^{-3}W\left( \frac{{\mathbf x}}{R}\right) 
\end{equation}
with a suitable dimensionless function \( W \). The normalization condition
becomes
\begin{equation}
\label{eq:norm-cond}
4\pi \int W\left( q\right) q^{2}dq=1.
\end{equation}

By construction, the window profile \( W\left( q\right)  \) should fall off
at \( q\sim 1 \) which corresponds to distances \( x \) of order \( H^{-1} \).
From Eq.~(\ref{eq:norm-cond}) it follows that \( W\left( q\right)  \) must
decay at least as \( q^{-4} \) at large \( q \); however, we shall see below
that the \( \propto q^{-4} \) or even \( \propto q^{-5} \) decay of the window
function introduces too much correlation between far-away points, and so we
shall assume that \( W\left( q\right)  \) falls off at least as \( q^{-6} \)
or faster at large \( q \). 

Smoothing is more conveniently performed directly in the Fourier domain, where
it corresponds to suppressing high-frequency modes of the field. The modes are
attenuated by the Fourier image of the window function. From the form (\ref{eq:wprof})
of the window function it follows that its Fourier image is a real function
of \( kR\left( t\right) \equiv p \):
\begin{equation}
\int e^{-i{\mathbf kx}}\frac{1}{R^{3}}W\left( \frac{x}{R}\right) d^{3}{\mathbf x}\equiv w\left( kR\right) ,
\end{equation}
 where
\begin{equation}
\label{eq:g-def}
w\left( p\right) =4\pi \int _{0}^{\infty }\frac{\sin pq}{p}W\left( q\right) qdq.
\end{equation}
 Since the profile \( W\left( q\right)  \) starts to decay at \( q\sim 1 \),
its Fourier image \( w\left( p\right)  \) becomes negligible at \( p\gg 1 \).
The mode expansion of the smoothed field is
\begin{equation}
\label{eq:sm-me}
\bar{\phi }\left( {\mathbf x},t\right) =\int \frac{d^{3}{\mathbf k}}{\left( 2\pi \right) ^{3/2}}w\left( kR\right) \psi _{k}\left( t\right) a_{{\mathbf k}}e^{i{\mathbf kx}}+h.c.
\end{equation}
 For instance, a Gaussian smoothing window in real space is
\begin{equation}
W_{s}\left( {\mathbf x};R\right) =\frac{1}{\left( 2\pi \right) ^{3/2}R^{3}}\exp \left[ -\frac{\left| {\mathbf x}\right| ^{2}}{2R^{2}}\right] 
\end{equation}
and in the Fourier domain
\begin{equation}
w\left( kR\right) =\exp \left( -\frac{k^{2}R^{2}}{2}\right) =\exp \left( -\frac{k^{2}\eta ^{2}}{2\epsilon ^{2}}\right) .
\end{equation}

Restrictions on the window profile \( W\left( q\right)  \) lead to conditions
on its Fourier image \( w\left( p\right)  \). Normalization of the window profile
means that \( w\left( 0\right) =1 \), while the fast decay of \( W\left( q\right)  \)
at large \( q \) makes \( w\left( p\right)  \) a differentiable function,
\begin{equation}
\left| \frac{d^{n}w\left( p\right) }{dp^{n}}\right| =\left| 4\pi \int _{0}^{\infty }\left[ \frac{d^{n}}{dp^{n}}\frac{\sin pq}{p}\right] qW\left( q\right) dq\right| <\infty \quad \textrm{if}\, p\neq 0,\, \left| W\left( q\right) \right| <Cq^{-n-2}\, \textrm{at}\, q\rightarrow \infty ,
\end{equation}
 
\begin{equation}
\left| \frac{d^{n}w\left( p\right) }{dp^{n}}\right| _{p=0}<\infty \quad \textrm{if}\, \left| W\left( q\right) \right| <Cq^{-n-4}\, \textrm{at}\, q\rightarrow \infty ,
\end{equation}
where \( C \) is a suitable constant. The odd derivatives of \( w\left( p\right)  \)
at \( p=0 \) generally vanish: 
\begin{equation}
\frac{d^{2n+1}w\left( 0\right) }{dp^{2n+1}}=4\pi \int _{0}^{\infty }\left[ \frac{d^{2n+1}}{dp^{2n+1}}\frac{\sin pq}{p}\right] _{p=0}qW\left( q\right) dq=0.
\end{equation}
(The latter condition is satisfied only if \( W\left( q\right)  \) falls off
faster than \( q^{-2n-4} \) at large \( q \) because only in that case it
is legitimate to take the limit of \( p=0 \) within the integral.) We find
that if \( W\left( q\right)  \) is decaying faster than \( q^{-4} \) viz.
\( q^{-5} \) then \( w^{\prime }\left( 0\right) =0 \) and \( w^{\prime \prime }\left( 0\right)  \)
is finite. We shall see shortly that the first two derivatives of \( w\left( p\right)  \)
at \( p=0 \) and smoothness of \( w\left( p\right)  \) at \( p>0 \) determine
the asymptotic behavior of the correlators for the smoothed field.

A consequence of the positivity of \( W\left( q\right)  \) is the non-vanishing
of the even derivatives of \( w\left( p\right)  \) at \( p=0 \) (provided
they exist), for instance
\begin{equation}
\label{eq:gsd}
w^{\prime \prime }\left( 0\right) =-\frac{4\pi }{3}\int _{0}^{\infty }q^{4}W\left( q\right) dq<0.
\end{equation}

If \( W\left( q\right)  \) falls off exponentially at \( q\rightarrow \infty  \),
the function \( w\left( p\right)  \) can be expanded at small \( p \) as
\begin{equation}
\label{eq:gp-exp}
w\left( p\right) =1+\frac{w^{\prime \prime }\left( 0\right) }{2!}p^{2}+\frac{w^{\left( 4\right) }\left( 0\right) }{4!}p^{4}+...,\quad w^{\left( 2n\right) }\left( 0\right) =\left( -\right) ^{n}\frac{4\pi }{2n+1}\int _{0}^{\infty }q^{2n+2}W\left( q\right) dq.
\end{equation}
The derivatives \( w^{\left( 2n\right) }\left( 0\right)  \) depend on the window
function but are generically of order \( 1 \) since \( W\left( q\right)  \)
falls off at \( q\sim 1 \) by construction. Our results depend only on the
first two terms of this series and on the differentiability of \( w\left( p\right)  \)
at all \( p \), so the condition \( W\left( q\right) \geq 0 \) can be dropped
as long as \( w^{\prime \prime }\left( 0\right) <0 \).

\section{Correlator of noise}

\label{sec:app-corr}Here we derive the correlators of the effective noise field
\( \xi \left( {\mathbf x},t\right)  \) for an arbitrary smoothing window.

The field \( \phi  \) is quantized using the mode expansion 
\begin{equation}
\label{eq:phime}
\phi \left( {\mathbf x},t\right) =\int \frac{d^{3}{\mathbf k}}{\left( 2\pi \right) ^{3/2}}\left( a_{{\mathbf k}}\psi _{k}\left( t\right) e^{i{\mathbf kx}}+h.c.\right) .
\end{equation}
The mode functions are
\begin{equation}
\label{eq:phimf}
\psi _{k}\left( \eta \right) =-\frac{He^{-ik\eta }}{\sqrt{2k}}\left( \eta +\frac{1}{ik}\right) 
\end{equation}
where the conformal time \( \eta  \) is defined as \( \eta \equiv -H^{-1}e^{-Ht} \),
with \( \partial _{t}=-H\eta \partial _{\eta } \). The mode functions \( \psi _{k}\left( \eta \right)  \)
satisfy the equation of motion, i.e. 
\begin{equation}
\left[ \frac{\partial ^{2}}{\partial t^{2}}+3H\frac{\partial }{\partial t}+k^{2}H^{2}\eta ^{2}\right] \psi _{k}\left( \eta \right) =0.
\end{equation}
We have defined the noise field \( \xi \left( {\mathbf x},t\right)  \) through
the time derivative of the averaged field \( \dot{\bar{\phi }} \) which has
a mode expansion 

\begin{equation}
\label{eq:noise-me}
\dot{\bar{\phi }}\left( {\mathbf x},t\right) =\int \frac{d^{3}{\mathbf k}}{\left( 2\pi \right) ^{3/2}}v_{k}\left( \eta \right) a_{{\mathbf k}}e^{i{\mathbf kx}}+h.c.,
\end{equation}
where
\begin{equation}
v_{k}\left( \eta \right) \equiv H\frac{d}{dt}\left[ w\left( kR\right) \psi _{k}\left( \eta \right) \right] =H\left[ -kRw^{\prime }\left( kR\right) \psi _{k}\left( \eta \right) +w\left( kR\right) \dot{\psi }_{k}\left( \eta \right) \right] .
\end{equation}
 In the limit of \( \epsilon \ll 1 \) we may disregard the second term in the
square brackets. The noise correlator is then

\begin{equation}
\label{eq:cdgen}
\left\langle \xi \left( {\mathbf x}_{1},\eta _{1}\right) \xi \left( {\mathbf x}_{2},\eta _{2}\right) \right\rangle =\frac{H^{4}\eta _{1}\eta _{2}}{4\pi ^{2}r\epsilon ^{2}}\int _{0}^{\infty }dk\, \sin kr\, h\left( k\right) ,
\end{equation}
\begin{equation}
h\left( k\right) \equiv \left( 1+ik\eta _{1}\right) \left( 1-ik\eta _{2}\right) e^{ik\left( \eta _{2}-\eta _{1}\right) }w^{\prime }\left( -\frac{k\eta _{1}}{\epsilon }\right) w^{\prime }\left( -\frac{k\eta _{2}}{\epsilon }\right) .
\end{equation}

From Eq.~(\ref{eq:cdgen}) we can obtain the asymptotic form of the correlator
at large effective distances \( \rho \gg 1 \). First we note that Eq.~(\ref{eq:cdgen})
is a sine transform of a certain (dimensionless) function which we denoted \( h\left( k\right)  \).
It is known that the asymptotic of a Fourier transform is determined by the
degree of smoothness of the transformed function. In the particular case of
the sine transform, repeated partial integration gives 
\begin{equation}
\label{eq:sin-trans}
\int _{0}^{\infty }h\left( k\right) \sin rk\, dk=\frac{h\left( 0\right) }{r}-\frac{h^{\prime \prime }\left( 0\right) }{r^{3}}+\frac{1}{r^{3}}\int _{0}^{\infty }dk\, \cos kr\, h^{\prime \prime \prime }\left( k\right) .
\end{equation}
If \( h^{\prime \prime \prime }\left( 0\right)  \) is finite, the last integral
is \( O\left( r^{-5}\right)  \) and Eq.~(\ref{eq:sin-trans}) may be used as
the asymptotic at large \( r \). Therefore we need to compute the second derivative
at \( k=0 \) of the relevant function \( h\left( k\right)  \) in our case,
and we can use the expansion (\ref{eq:gp-exp}) for the window function \( w\left( p\right)  \).
We also find that the asymptotic expansion of Eq.~(\ref{eq:sin-trans}) is valid,
i.e.\ \( h^{\prime \prime \prime }\left( 0\right)  \) is finite, if
\begin{equation}
\label{eq:g-cond}
w^{\prime }\left( 0\right) =0,\quad \left| w^{\prime \prime }\left( 0\right) \right| <\infty .
\end{equation}
 As we have seen above, these conditions are satisfied if for example \( \left| W\left( q\right) \right| <Cq^{-6} \)
at \( q\rightarrow \infty  \). The asymptotic form of Eq.~(\ref{eq:cdgen})
at large \( r \) is then
\begin{equation}
\label{eq:smcor-asy}
\left\langle \xi \left( {\mathbf x}_{1},\eta _{1}\right) \xi \left( {\mathbf x}_{2},\eta _{2}\right) \right\rangle =-\frac{\left( H^{2}\eta _{1}\eta _{2}\right) ^{2}}{2\pi ^{2}r^{4}\epsilon ^{4}}\left| w^{\prime \prime }\left( 0\right) \right| ^{2}+O\left( r^{-6}\right) .
\end{equation}
This expression is very similar to Eq.~(\ref{eq:shcor-asy}) for the unsmoothed
correlator of field derivatives. We note that the asymptotic Eq.~(\ref{eq:smcor-asy})
is essentially independent of the shape of the window function, since the value
\( \left| w^{\prime \prime }\left( 0\right) \right|  \) as indicated by Eq.~(\ref{eq:gsd})
has the meaning of the window-averaged squared distance and must be of order
\( 1 \) because the window profile \( W\left( q\right)  \) starts to decay
at \( q\sim 1 \) by construction. 

We can obtain a simpler expression for the correlator in the limit when the
smoothing parameter \( \epsilon  \) is small while the product \( \epsilon Hr \)
is finite. A rescaling \( r\rightarrow \epsilon Hr\equiv \rho  \) and the corresponding
change of variable \( k\equiv \epsilon H\kappa  \) simplify Eq.~(\ref{eq:cdgen})
because we can omit terms of order \( \epsilon  \) and smaller; in particular,
the product of mode functions is simplified to
\begin{equation}
\psi ^{*}_{k}\left( \eta _{1}\right) \psi _{k}\left( \eta _{2}\right) =\frac{1}{2H\kappa ^{3}\epsilon ^{3}}\left( 1+O\left( \epsilon ^{2}\right) \right) .
\end{equation}
 The leading term in the correlator, expressed through \( \kappa  \) and \( \rho  \),
becomes
\begin{equation}
\label{eq:smcorgen}
\left\langle \xi \left( {\mathbf x}_{1},\eta _{1}\right) \xi \left( {\mathbf x}_{2},\eta _{2}\right) \right\rangle =\frac{H^{6}\eta _{1}\eta _{2}}{4\pi ^{2}\rho }\int _{0}^{\infty }d\kappa \, \sin \kappa \rho \, w^{\prime }\left( -H\eta _{1}\kappa \right) w^{\prime }\left( -H\eta _{2}\kappa \right) +O\left( \epsilon ^{2}\right) .
\end{equation}
 We find that in the limit of small \( \epsilon  \) but finite \( \epsilon Hr \)
the correlator as function of the ``effective distance'' \( \rho  \) and the
time difference (expressed by \( \eta _{2}/\eta _{1} \)) becomes independent
of \( \epsilon  \).

Eq.~(\ref{eq:smcorgen}) allows us to compute the correlator at all distances
in the limit of small \( \epsilon  \). For a Gaussian smoothing window, \( w\left( p\right) =\exp \left( -p^{2}/2\right)  \),
we obtain in the limit of small \( \epsilon  \) 
\[
\left\langle \xi \left( {\mathbf x}_{1},\eta _{1}\right) \xi \left( {\mathbf x}_{2},\eta _{2}\right) \right\rangle =\frac{\left( H^{4}\eta _{1}\eta _{2}\right) ^{2}}{4\pi ^{2}\rho }\int _{0}^{\infty }\exp \left[ -H^{2}\frac{\eta _{1}^{2}+\eta _{2}^{2}}{2}\kappa ^{2}\right] \kappa ^{2}\sin \kappa \rho d\kappa \]
\begin{equation}
\label{eq:smcor-gau}
=\frac{\left( H^{4}\eta _{1}\eta _{2}\right) ^{2}\nu ^{4}}{4\pi ^{2}\rho ^{4}}\left[ 1-\left( \frac{1}{\nu }-\nu \right) i\sqrt{\frac{\pi }{2}}\textrm{erf}\left( \frac{i\nu }{\sqrt{2}}\right) \exp \left( -\frac{\nu ^{2}}{2}\right) \right] ,
\end{equation}
where we introduced a dimensionless quantity
\begin{equation}
\nu \equiv \frac{\rho }{\sqrt{H^{2}\left( \eta _{1}^{2}+\eta _{2}^{2}\right) }}.
\end{equation}
The shape of this function for \( \eta _{1}=\eta _{2} \) is shown in Fig.~1.
The leading term of the expression in brackets in Eq.~(\ref{eq:smcor-gau})
at large \( \nu  \) is \( \left( -2\nu ^{-4}\right)  \), and since for the
Gaussian window \( w^{\prime \prime }\left( 0\right) =-1 \), we recover Eq.~(\ref{eq:smcor-asy}).
The value of the correlator at coincident points (\( \rho =0 \)) as function
of time separation is (cf.~Eq.~(\ref{eq:shcor-asy-t}))
\begin{equation}
\left\langle \xi \left( 0,\eta _{1}\right) \xi \left( 0,\eta _{2}\right) \right\rangle =\frac{\left( H^{4}\eta _{1}\eta _{2}\right) ^{2}}{2\pi ^{2}\left( \eta _{1}^{2}+\eta _{2}^{2}\right) ^{2}}=\frac{H^{4}}{8\pi ^{2}}\frac{1}{\cosh ^{2}H\Delta t}.
\end{equation}

One can also obtain the leading asymptotics of the unequal-time correlator at
large time separations. We again start with Eq.~(\ref{eq:cdgen}) and assume
that the time separation is much greater than the Hubble time, \( \eta _{2}/\eta _{1}\equiv a^{-1}\ll 1 \).
For simplicity we can take \( H\eta _{1}=-1 \). We use Eq.~(\ref{eq:gp-exp})
for \( w\left( a^{-1}k\right)  \) at small \( a^{-1}\kappa  \) (since the
integration is effectively performed over a fixed finite range of \( k \))
to obtain

\begin{equation}
\label{eq:cor-asy-t}
\left\langle \xi \left( {\mathbf x}_{1},\eta _{1}\right) \xi \left( {\mathbf x}_{2},\eta _{2}\right) \right\rangle =\frac{H^{2}w^{\prime \prime }\left( 0\right) }{4\pi ^{2}\epsilon ^{3}a^{2}Hr}\int _{0}^{\infty }dk\, k\sin kr\, w^{\prime }\left( \frac{k}{\epsilon H}\right) e^{ik/H}\left( 1+i\frac{k}{H}\right) +O\left( a^{-4}\right) .
\end{equation}
The integral in Eq.~(\ref{eq:cor-asy-t}) is clearly time-independent. Therefore
the correlator decays as \( a^{-2}=\exp \left( -2Ht\right)  \) with time separation
at any fixed distance.

\section{Correlators of unsmoothed fields}

\label{sec:app-desitter}The correlator of time derivatives of the field at
equal times is
\begin{equation}
\left\langle \dot{\phi }\left( {\mathbf x},t\right) \dot{\phi }\left( {\mathbf x}^{\prime },t\right) \right\rangle =\frac{1}{\left( 2\pi \right) ^{3}}\int \left| \dot{\psi }_{k}\left( t\right) \right| ^{2}e^{i{\mathbf k}\left( {\mathbf x}-{\mathbf x}^{\prime }\right) }d^{3}{\mathbf k}=\frac{1}{2\pi ^{2}}\int _{0}^{\infty }\frac{\sin kr}{r}\left| \dot{\psi }_{k}\left( t\right) \right| ^{2}kdk,
\end{equation}
where \( r\equiv \left| {\mathbf x}-{\mathbf x}^{\prime }\right|  \) is the distance
between points. (In this Appendix, angular brackets denote quantum expectation
values.) The derivatives of the mode function are
\begin{equation}
\frac{\partial \psi _{k}\left( \eta \right) }{\partial \eta }=\frac{ikH\eta }{\sqrt{2k}}e^{-ik\eta },\quad \dot{\psi }_{k}\equiv \frac{\partial \psi _{k}}{\partial t}=-\frac{ik\left( H\eta \right) ^{2}}{\sqrt{2k}}e^{-ik\eta },
\end{equation}
which gives (after a regularization of the integral) at \( t=0 \) (\( H\eta =-1 \))
\begin{equation}
\left\langle \dot{\phi }\left( {\mathbf x},t\right) \dot{\phi }\left( {\mathbf x}^{\prime },t\right) \right\rangle =\frac{1}{4\pi ^{2}}\int _{0}^{\infty }\frac{\sin kr}{r}k^{2}dk=\frac{1}{4\pi ^{2}}\frac{d^{2}}{dr^{2}}\int _{0}^{\infty }\sin kr\, dk=-\frac{1}{2\pi ^{2}r^{4}}.
\end{equation}

The unequal-time correlator at coincident points (\( r=0 \)) is found as
\[
\left\langle \dot{\phi }\left( 0,\eta _{1}\right) \dot{\phi }\left( 0,\eta _{2}\right) \right\rangle =\frac{1}{2\pi ^{2}}\int _{0}^{\infty }\dot{\psi }^{*}_{k}\left( \eta _{1}\right) \dot{\psi }_{k}\left( \eta _{2}\right) k^{2}dk
=\frac{\left( H^{2}\eta _{1}\eta _{2}\right) ^{2}}{4\pi ^{2}}\int _{0}^{\infty }k^{3}e^{-ik\left( \eta _{2}-\eta _{1}\right) }dk
\]
\begin{equation}
\label{eq:shcor-asy-t}
=\frac{1}{4\pi ^{2}}i^{3}\frac{d^{3}}{d\eta _{2}^{3}}\int _{0}^{\infty }e^{-ik\left( \eta _{2}-\eta _{1}\right) }dk=\frac{6H^{4}}{4\pi ^{2}}\frac{\eta _{1}^{2}\eta _{2}^{2}}{\left( \eta _{1}-\eta _{2}\right) ^{4}}=\frac{3H^{4}}{32\pi ^{2}}\left( \sinh \frac{H\left| t_{1}-t_{2}\right| }{2}\right) ^{-4}.
\end{equation}

Finally, we consider the general correlator at arbitrary points and times:
\[
\left\langle \dot{\phi }\left( {\mathbf x}_{1},t_{1}\right) \dot{\phi }\left( {\mathbf x}_{2},t_{2}\right) \right\rangle =\frac{1}{2\pi ^{2}}\int _{0}^{\infty }\dot{\psi }_{k}\left( t\right) \dot{\psi }_{k}^{*}\left( 0\right) \frac{\sin kr}{r}kdk
\]
\begin{equation}
\label{eq:shcor-gen}
=\frac{H^{4}}{2\pi ^{2}}\left( \eta _{1}\eta _{2}\right) ^{2}\frac{3\left( \eta _{1}-\eta _{2}\right) ^{2}+r^{2}}{\left( \left( \eta _{1}-\eta _{2}\right) ^{2}-r^{2}\right) ^{3}}.
\end{equation}
As expected, it diverges on the lightcone \( r=\left| \eta _{1}-\eta _{2}\right|  \).
The asymptotic form of Eq.~(\ref{eq:shcor-gen}) at large distances \( r \)
is
\begin{equation}
\label{eq:shcor-asy}
\left\langle \dot{\phi }\left( {\mathbf x}_{1},t_{1}\right) \dot{\phi }\left( {\mathbf x}_{2},t_{2}\right) \right\rangle =-\frac{H^{4}\left( \eta _{1}\eta _{2}\right) ^{2}}{2\pi ^{2}r^{4}}+O\left( r^{-6}\right) .
\end{equation}

\end{document}